\documentclass[fleqn,11pt,twoside]{article}
\usepackage{espcrc2}
\usepackage[ruled]{algorithm2e}
\usepackage{latexsym,amssymb,amsmath,amssymb,xspace,theorem}
\usepackage{multicol}
\usepackage{t1enc}
\usepackage{epic,eepic}
\usepackage[final]{graphicx}
\usepackage{xspace}
\usepackage[latin1]{inputenc}
\usepackage[T1]{fontenc}
\newcommand{\ignore}[1]{}
\newcommand\ligne{\par $\;$\par}

\newtheorem{theoreme}{Theorem}[section]

\newtheorem{jmclaim}{Claim}[theoreme]

\newtheorem{definition}[theoreme]{Définition}

\newenvironment{preuve}{{\bf \noindent Proof } }{\hfill$\square$\\}
\newenvironment{preuvedeclaim}{{\bf Proof }}{{\hfill\tiny{$\blacksquare$\\}}}

\title{On $(P_5,\overline{P_5})$-sparse graphs and other families}

\author{Jean-Luc \textsc{Fouquet}
and Jean-Marie \textsc{Vanherpe}\\
email~: \{Jean-Luc.Fouquet, Jean-Marie.Vanherpe\}@univ-orleans.fr}

\begin{document}

%\section*{Abstract}
\begin{abstract}
We extend the notion of  $P_4$-sparse graphs previously introduced
by {\scshape Hoàng} in \cite{Hoa85} by considering
$\mathcal{F}$-sparse graphs were $\mathcal{F}$ denotes a finite set
of graphs on $p$ vertices. Thus we obtain some results on
$(P_5,\overline{P_5})$-sparse graphs  already  known on
$(P_5,\overline{P_5})$-free graphs. Finally we completely describe
the structure of $(P_5,\overline{P_5}, bull$)-sparse graphs, it
follows that those graphs have bounded clique-width.
\end{abstract}

\maketitle
\section{Introduction}
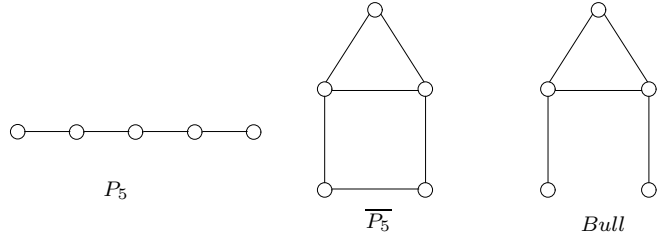
\begin{figure*}[!ht]
\begin{center}
\setlength{\unitlength}{0.00017498in}
\begingroup\makeatletter\ifx\SetFigFont\undefined
% extract first six characters in \fmtname
\def\x#1#2#3#4#5#6#7\relax{\def\x{#1#2#3#4#5#6}}%
\expandafter\x\fmtname xxxxxx\relax \def\y{splain}%
\ifx\x\y   % LaTeX or SliTeX?
\gdef\SetFigFont#1#2#3{%
  \ifnum #1<17\tiny\else \ifnum #1<20\small\else
  \ifnum #1<24\normalsize\else \ifnum #1<29\large\else
  \ifnum #1<34\Large\else \ifnum #1<41\LARGE\else
     \huge\fi\fi\fi\fi\fi\fi
  \csname #3\endcsname}%
\else
\gdef\SetFigFont#1#2#3{\begingroup
  \count@#1\relax \ifnum 25<\count@\count@25\fi
  \def\x{\endgroup\@setsize\SetFigFont{#2pt}}%
  \expandafter\x
    \csname \romannumeral\the\count@ pt\expandafter\endcsname
    \csname @\romannumeral\the\count@ pt\endcsname
  \csname #3\endcsname}%
\fi
\fi\endgroup
{\renewcommand{\dashlinestretch}{30}
\begin{picture}(20169,7367)(0,-10)
\put(3301,1294){\makebox(0,0)[b]{\smash{{{\SetFigFont{8}{9.6}{rm}$P_5$}}}}}
\path(1861,3319)(466,3319)
\put(3912,3309){\ellipse{450}{450}}
\path(3706,3319)(2311,3319)
\put(5757,3309){\ellipse{450}{450}}
\path(5551,3319)(4156,3319)
\put(7602,3309){\ellipse{450}{450}}
\path(7396,3319)(6001,3319)
\put(16786,1494){\ellipse{450}{450}}
\put(19936,1494){\ellipse{450}{450}}
\put(19936,4644){\ellipse{450}{450}}
\put(16786,4644){\ellipse{450}{450}}
\put(18361,7119){\ellipse{450}{450}}
\put(11386,7119){\ellipse{450}{450}}
\put(9811,4644){\ellipse{450}{450}}
\put(12961,4644){\ellipse{450}{450}}
\put(12961,1494){\ellipse{450}{450}}
\put(9811,1494){\ellipse{450}{450}}
\put(233,3319){\ellipse{450}{450}}
\path(10051,1519)(12751,1519)
\path(19891,4869)(18541,6984)
\path(16831,4869)(18181,6984)
\path(19936,4419)(19936,1719)
\path(16786,4419)(16786,1719)
\path(17011,4599)(19711,4599)
\path(10036,4599)(12736,4599)
\path(9811,4419)(9811,1719)
\path(12961,4419)(12961,1719)
\path(9856,4869)(11206,6984)
\path(12916,4869)(11566,6984)
\put(18511,214){\makebox(0,0)[b]{\smash{{{\SetFigFont{8}{9.6}{rm}$Bull$}}}}}
\put(11491,304){\makebox(0,0)[b]{\smash{{{\SetFigFont{8}{9.6}{rm}$\overline{P_5}$}}}}}
\put(2067,3309){\ellipse{450}{450}}
\end{picture}
}
\caption{The forbidden configurations in a $(P_5,\overline{P_5},Bull)$-free graph}\label{fig:P5HB}
\end{center}
\end{figure*}

$P_4$-free graphs, also called {\em Cographs}, were designed to be completely decomposable by complementation and
motivated researchers for studying graph classes characterized with forbidden configurations. In addition, a number of optimization problems on a graph can be reduced to their weighted version on the set of prime graphs also called the set of representative graphs (recall that the representative graph of graph $G$ is obtained from $G$ by contracting
every maximal proper module of $G$ into a single vertex)(see \cite{MohRad84}). Thus sub-classes of $P_5$- free graphs
were intensively studied (see e.g. \cite{BacTuz90,BraKra05,BraMos03}), in particular  Fouquet in \cite{Fou93}
consider $(P_5,\overline{P_5})$-free graphs and the subclass of $(P_5,\overline{P_5},Bull)$-free graphs (see Figure
\ref{fig:P5HB}). Later Giakoumakis and Rusu \cite{GiaRus97} provide efficient solutions for some
optimization problems on $(P_5,\overline{P_5})$-free graphs.

Hoàng introduced in \cite{Hoa85} the $P_4$-sparse graphs (every induced subgraph on $5$ vertices contains at most one $P_4$) and several extensions of this notion have arisen in the litterature (see
 for examples \cite{BabOla96,FouVan07,GiaRouThu97b,RouRusThu99}).
We are concerned here with ($P_5$,$\overline{P_5}$)-graphs and ($P_5$,$\overline{P_5}$,$Bull$)-graphs where these classes of graphs are defined in the same way (every subgraph on $6$ vertices contains at most one subgraph in the family).

In this paper, we extend the notion of $P_4$-sparse in the following way~: A graph $G$ is said to be
$\mathcal{F}$-sparse, where $\mathcal{F}$ denotes a set of graphs of order $p$, whenever any induced subgraph of $G$ on
$p+1$ vertices contains at most one graph of $\mathcal{F}$ as induced subgraph. \ignore{We first propose a theorem which
recursively reduces the recognition problem of $\mathcal{F}$-sparse graphs (when $\mathcal{F}$ is a set of prime
graphs) to those of the set of  representative graphs (recall that the representative graph of graph $G$ is obtained
from $G$ by contracting every maximal proper module of $G$ into a single vertex).} We study
$\mathcal{F}$-sparse when $\mathcal{F}=\{P_5,\overline{P_5}\}$) and when $\mathcal{F}=\{P_5,\overline{P_5}, Bull\}$). Those graphs classes are defined with configurations which
are prime with respect to modular decomposition (see Figure \ref{fig:P5HB}) and which properly intersect graphs classes
such that $PL$-graphs or some $(q,t)$-graphs classes.

We obtain some results on $(P_5,\overline{P_5})$-sparse graphs  already  known on
$(P_5,\overline{P_5})$-free graphs and we completely describe
the structure of $(P_5,\overline{P_5}, bull$)-sparse graphs. This
shows  that those graphs have bounded clique-width.

\subsection*{Basics}
Let $G=(V,E)$ be a graph, the complementary graph of $G$ is denoted
$\overline{G}$. If $x$ and $y$ are two adjacent vertices of $G$, $x$
is said adjacent to $y$ and  $y$ is a {\em neighbor} of $x$. A graph
on $2n$ vertices such that all of them have exactly one neighbor is
a $nK_2$.

Let $X$ be a set of vertices and $x$  be a vertex such that $x\notin
X$, the set of neighbors of $x$ that belong to $X$ is said the {\em
neighborhood } of $x$ in $X$ and is denoted $N_X(x)$, if
$N_X(x)=\emptyset$ $x$ is said {\em independent} of $X$ and {\em
total} for $X$ when $N_X(x)=X$, if $x$ is not independent of $X$ nor
total for $X$, $x$ is said {\em partial} for $X$. If $x$ is
independent of $X$ (resp. total for $X$), $x$ is said isolated in
$X\cup\{x\}$ (universal for $X\cup\{x\}$).

Let $X$ and $Y$ be two disjoint sets of vertices, the set
$\underset{y\in Y}{\bigcup} N_X(y)$ is denoted $N_X(Y)$ and called
the {\em neighborhood of $Y$ in $X$}. If there is no edge connecting
a vertex of $X$ to a vertex of $Y$, the sets $X$ and $Y$ are
independent while $X$ is total for $Y$ when there is all possible
edges connecting vertices of $X$ to vertices of $Y$.

%ignore
\section{On (\ensuremath{P_5,\overline{P_5}})-sparse graphs.}
In this section we consider $\mathcal{F}$-sparse graphs when $\mathcal{F}=\{P_5,\overline{P_5}\}$ and we call those graphs  $(P_5,\overline{P_5})$-sparse. Recall that in a such graph every induced subgraph on $6$ vertices contains at most one $P_5$ or $\overline{P_5}$.
\begin{theoreme}\label{thm:LesC5DansLesP5H-Sparse}
A prime $(P_5,\overline{P_5})$-sparse graph is either $C_5$-free or isomorphic to a $C_5$.
\end{theoreme}
\begin{preuve}
Let $G$ be a prime $\mathcal{F}$-sparse graph having at least $6$ vertices.

Observe first that a vertex, say $x$, which is partial to a $C_5$ of $G$ is either adjacent  to exactly two
non-adjacent vertices of the $C_5$ or to three consecutive vertices of the $C_5$. In all other cases of adjacencies the
subgraph induced by the vertices of the $C_5$ together  witrh $x$ contains two $P_5$ or $\overline{P_5}$, a
contradiction.

Let $abcde$ be a $C_5$ of $G$, since $G$ is prime there must exist in $G$ a vertex, say $x$ which is partial to $abcde$. Without loss of generality we can assume that $x$ is adjacent to $a$ and $c$ and independent of $d$ and $e$. Let $A$ be the set of vertices of $G$ which are adjacent to $a$ and $c$ and independent of $d$ and $e$. Since $A$ contains at least two vertices ($\{b,x\}\subseteq A$) and $G$ is prime there must be a vertex, say $y$, outside of $A$ which distinguishes two vertices of $A$ say $b_1$ and $b_2$. But now, the vertex $y$ cannot be outside of $A$ and satisfy the above observation with bots $C_5$ $ab_1cde$ and $ab_2cde$, a contradiction.
\end{preuve}
\ignore{
\begin{figure*}[!ht]
\begin{center}
\input{./Figures/WelshPowellPerfect.eepic}
\caption{The $17$ forbidden configurations for a Welsh-Powell perfect graph.}\label{fig:WelshPowellPerfect}
\end{center}
\end{figure*}
}
 Welsh-Powell perfect graphs are perfectly orderable and are
characterized with $17$ forbidden  configurations (see
\cite{ChvHoaMahWer87}). It is a straightforward exercise to see that
$(P_5,\overline{P_5})$-sparse graphs which are also $C_5$-free are
Welsh-Powell perfect \ignore{(see Figure
\ref{fig:WelshPowellPerfect})}. In \cite{Hoa93}, Hoàng, gives
algorithms to solve the {\em Maximum Weighted Clique} problem as
well as the {\em Minimum Weighted Coloring} problem on perfectly
orderable graphs within $O(nm)$ time complexity.

Thus, as well as for $(P_5,\overline{P_5})$-free graphs (see \cite{GiaRus97}),  there exists algorithms running in
$O(nm)$ time, for computing a {\em Maximum Weigted Clique} and a {\em Minimum Weighted Coloring} in a weighted $(P_5,
\overline{P_5})$-sparse graph.\ignore{(see Algorithm \ref{alg:OptimizationOnP5HSparse})} Since the class of
$(P_5,\overline{P_5})$-sparse graphs is auto-complementary the parameters  {\em Maximum Weighted Stable Set} and {\em
Minimum Weighted Clique Cover} can be computed within the same time complexity.

\ignore{
%\nocaptionofalgo
\dontprintsemicolon
\Setnlsty{textbf}{Step~}{~:}
\begin{algorithm*}[!ht]\label{alg:OptimizationOnP5HSparse}
\caption{Maximum Weighted Clique (resp. Minimum Weighted Coloring)
for $(P_5,\overline{P_5})$-sparse graphs.}\KwIn{$(G,w)$ a weighted
$(P_5,\overline{P_5})$-sparse graph.} \KwOut{ A Maximum Weighted
Clique (resp. A minimum Weighted Coloring) of $G$.} \BlankLine
\lnl{alg:step1} Compute the modular decomposition tree of $G$ and
its set of prime representative graphs (see
\cite{ConSpi94,CouHab94,DahGusCon01})\; \lnl{alg:step2}
\lForEach{prime representative graph $H$ of $G$} {\;
    \Indp\eIf{$H$ is not a $C_5$}
        {\tcc{By Theorem \ref{thm:LesC5DansLesP5H-Sparse} $H$ is $C_5$-free}\;
        Use Hoàng's algorithm (\cite{Hoa93}) to compute a {\em Maximum Weighted Clique} (resp. a {\em Minimum Weighted Coloring}) }
    {Compute a {\em Maximum Weighted Clique} of $H$ (resp. {\em Minimum Weighted Coloring} of $H$, see \cite{Rus99}\ignore{\cite{CouGiaVan97}}) \;
    }
}
\lnl{alg:step3} Compute a {\em Maximum Weighted Clique} (resp. {\em Minimum Weighted Coloring} for $G$ by a bottom-up traversal of its modular decomposition tree (see \cite{MohRad84})\;
\end{algorithm*}
}%ignore
\section{(\ensuremath{P_5,\overline{P_5}, Bull})-sparse graphs.}
In this section we will study $\mathcal{F}$-sparse graphs where $\mathcal{F}=\{P_5,\overline{P_5}, Bull\}$, namely the $(P_5,\overline{P_5}, Bull)$-sparse graphs. We will characterize the prime graphs of this family and give some consequences.

Let's first recall a main result on $(P_5,\overline{P_5}, Bull)$-free graphs.
\begin{theoreme}\label{thm:StructureDesP5HB-Free}(\cite{Fou93})
A prime graph $G$ is $(P_5, \overline{P_5}, bull)$-free if and only if one of the following holds~:
\begin{enumerate}
 \item $G$ is isomorphic to a $C_5$
 \item $G$ or its complement is a bipartite $P_5$-free graph.
\end{enumerate}
\end{theoreme}

Since Theorem \ref{thm:LesC5DansLesP5H-Sparse} also holds for $(P_5,\overline{P_5}, Bull)$-sparse graphs  we consider
henceforth only $C_5$-free graphs.
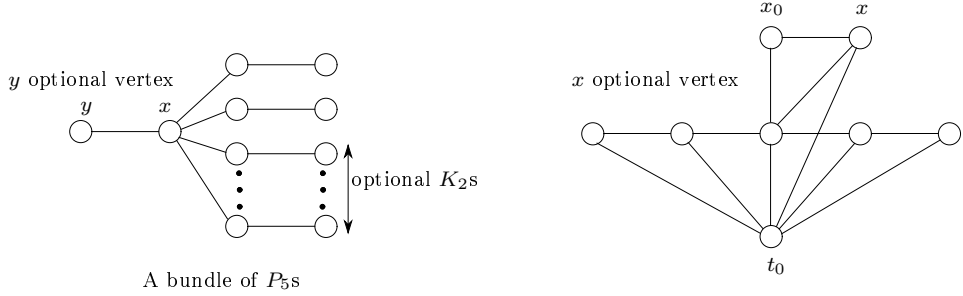
\begin{figure*}[!ht]
\begin{center}
  \setlength{\unitlength}{0.00026247in}
\begingroup\makeatletter\ifx\SetFigFont\undefined
% extract first six characters in \fmtname
\def\x#1#2#3#4#5#6#7\relax{\def\x{#1#2#3#4#5#6}}%
\expandafter\x\fmtname xxxxxx\relax \def\y{splain}%
\ifx\x\y   % LaTeX or SliTeX?
\gdef\SetFigFont#1#2#3{%
  \ifnum #1<17\tiny\else \ifnum #1<20\small\else
  \ifnum #1<24\normalsize\else \ifnum #1<29\large\else
  \ifnum #1<34\Large\else \ifnum #1<41\LARGE\else
     \huge\fi\fi\fi\fi\fi\fi
  \csname #3\endcsname}%
\else
\gdef\SetFigFont#1#2#3{\begingroup
  \count@#1\relax \ifnum 25<\count@\count@25\fi
  \def\x{\endgroup\@setsize\SetFigFont{#2pt}}%
  \expandafter\x
    \csname \romannumeral\the\count@ pt\expandafter\endcsname
    \csname @\romannumeral\the\count@ pt\endcsname
  \csname #3\endcsname}%
\fi
\fi\endgroup
{\renewcommand{\dashlinestretch}{30}
\begin{picture}(19566,6104)(0,-10)
\put(339,4198){\makebox(0,0)[lb]{\smash{{{\SetFigFont{8}{9.6}{rm}$y$ optional vertex}}}}}
\put(6749,2829){\ellipse{450}{450}}
\path(5174,2829)(6524,2829)
\put(6679,1763){\blacken\ellipse{96}{96}}
\put(6679,1763){\ellipse{96}{96}}
\put(6691,2104){\blacken\ellipse{96}{96}}
\put(6691,2104){\ellipse{96}{96}}
\put(6683,2439){\blacken\ellipse{96}{96}}
\put(6683,2439){\ellipse{96}{96}}
\put(5007,1767){\blacken\ellipse{96}{96}}
\put(5007,1767){\ellipse{96}{96}}
\put(5007,2104){\blacken\ellipse{96}{96}}
\put(5007,2104){\ellipse{96}{96}}
\put(4995,2439){\blacken\ellipse{96}{96}}
\put(4995,2439){\ellipse{96}{96}}
\put(6749,1367){\ellipse{450}{450}}
\put(4949,1367){\ellipse{450}{450}}
\path(5174,1367)(6524,1367)
\put(6749,4629){\ellipse{450}{450}}
\put(4949,4629){\ellipse{450}{450}}
\path(5174,4629)(6524,4629)
\put(12133,3239){\ellipse{450}{450}}
\put(13933,3239){\ellipse{450}{450}}
\path(12358,3239)(13708,3239)
\put(15733,3239){\ellipse{450}{450}}
\path(14158,3239)(15508,3239)
\put(17533,3239){\ellipse{450}{450}}
\path(15958,3239)(17308,3239)
\put(19333,3239){\ellipse{450}{450}}
\path(17758,3239)(19108,3239)
\put(17533,5180){\ellipse{450}{450}}
\path(15958,5180)(17308,5180)
\put(1800,3279){\ellipse{450}{450}}
\put(3599,3279){\ellipse{450}{450}}
\put(4949,3729){\ellipse{450}{450}}
\put(6749,3729){\ellipse{450}{450}}
\put(15733,5180){\ellipse{450}{450}}
\put(15733,1158){\ellipse{450}{450}}
\path(12219,3048)(15513,1226)
\path(14056,3062)(15570,1319)
\blacken\path(7268.000,2761.000)(7196.000,3001.000)(7124.000,2761.000)(7196.000,2833.000)(7268.000,2761.000)
\path(7196,3001)(7196,1233)
\blacken\path(7124.000,1473.000)(7196.000,1233.000)(7268.000,1473.000)(7196.000,1401.000)(7124.000,1473.000)
\path(3780,3161)(4732,2857)
\path(2025,3279)(3375,3279)
\path(3734,3459)(4814,4449)
\path(3824,3324)(4724,3684)
\path(3734,3099)(4764,1513)
\path(5174,3729)(6524,3729)
\path(15723,3022)(15723,1379)
\path(15723,3472)(15723,4963)
\path(15864,3416)(17383,5019)
\path(15842,1360)(17460,4973)
\path(15896,1310)(17425,3043)
\path(15945,1211)(19148,3107)
\put(3039,148){\makebox(0,0)[lb]{\smash{{{\SetFigFont{8}{9.6}{rm}A bundle of $P_5$s}}}}}
\put(15516,5693){\makebox(0,0)[lb]{\smash{{{\SetFigFont{8}{9.6}{rm}$x_0$ }}}}}
\put(15654,456){\makebox(0,0)[lb]{\smash{{{\SetFigFont{8}{9.6}{rm}$t_0$ }}}}}
\put(1790,3706){\makebox(0,0)[lb]{\smash{{{\SetFigFont{8}{9.6}{rm}$y$}}}}}
\put(3375,3684){\makebox(0,0)[lb]{\smash{{{\SetFigFont{8}{9.6}{rm}$x$}}}}}
\put(0,3279){\makebox(0,0)[lb]{\smash{{{\SetFigFont{5}{6.0}{rm} }}}}}
\put(1577,3247){\makebox(0,0)[lb]{\smash{{{\SetFigFont{5}{6.0}{rm}$\;$}}}}}
\put(17446,5653){\makebox(0,0)[lb]{\smash{{{\SetFigFont{8}{9.6}{rm}$x$ }}}}}
\put(7260,2175){\makebox(0,0)[lb]{\smash{{{\SetFigFont{8}{9.6}{rm}optional $K_2$s}}}}}
\put(11711,4209){\makebox(0,0)[lb]{\smash{{{\SetFigFont{8}{9.6}{rm}$x$ optional vertex}}}}}
\put(4949,2829){\ellipse{450}{450}}
\end{picture}
}
  \caption{The $2$ types of prime $(P_5,\overline{P_5},Bull)$-sparse graphs which are $C_5$-free and contain a $P_5$.}\label{fig:FaisceauDeP5}
  \end{center}
  \end{figure*}
\begin{theoreme}\label{thm:P5HBSparseAvecP5}
Let $G$ be a prime $C_5$-free which contains an induced $P_5$ (resp. $\overline{P_5}$). \\
$G$ is $(P_5,\overline{P_5},Bull)$-sparse if and only if $G$ (resp. $\overline{G}$) is isomorphic to one of the graphs depicted in Figure \ref{fig:FaisceauDeP5}.
\end{theoreme}

%ignore
\begin{preuve}{\bf of Theorem \ref{thm:P5HBSparseAvecP5}.} It is easy to see that the graphs depicted in Figure
\ref{fig:FaisceauDeP5} are prime $(P_5,\overline{P_5}, Bull)$-sparse graphs, consequently  in the following we consider
the only if part of the theorem.

 Assume without loss of generality that $G$ contains a $P_5$, namely $abcde$.  Observe first that a vertex partial to this $P_5$ can only be adjacent to $c$, all other adjacency cases lead to a contradiction
Let's denote $C$ the set of vertices in $G$ whose neighborhood in $\{a, b, c, d, e\}$ is $\{c\}$,  in addition we
denote $I$ the set of vertices of $G$ which have no neighbor in $\{a, b, c, d, e \}$ while $T$ denotes the set of
vertices of $G$ which are total for $\{a, b, c, d, e\}$, note that $V(G)=\{a, b, c, d, e\}\cup C \cup T \cup I$.
Moreover we suppose henceforth that $C$ is not empty, otherwise by the primality assumption, $G$ would be the $P_5$
$abcde$ itself (one of the graphs depicted in Figure \ref{fig:FaisceauDeP5}).
\begin{jmclaim}\label{thm:P5HBSparseAvecP5:Clm1}
If $I$ has a neighbor in $C$  then $N_C(I)\cup N_I(C)$ is a $nK_2$, the vertices of $N_I(C)$ are isolated in $I$ and
the vertices of $N_C(I)$ are isolated in $C$. Moreover $T$ is total for $C$ and $N_I(C)$.
\end{jmclaim}
\begin{preuvedeclaim}
Let's assume that $x\in C$ has a neighbor $i\in I$, so $\{a,b,c,x,i\}$ is a $P_5$. Then $x$ (resp. $i$)has no other
neighbor in $I$ (resp. $C$). Moreover $N_I(C)$ is isolated in $I$ because if $i$ has a neighbor $i'$ in $I$, then
$\{a,b,c,x,i,i'\}$ is a $P_6$, a contradiction; and $N_C(I)$ is isolated in $C$ because if $x$ has a neighbor $x'$ in
$C$ then $\{a,b,c,x,x',i$ induces a $P_5$ and a $bull$, a contradiction. Let $t\in T$, assume that $t$ isn't a neighbor
of $x$ or $i$. Let first $t$ isn't a neighbor of $i$ then $\{i,x,c,t,a,e\}$ induces $2$ $P_5$ or $2$ $bull$. Otherwise
if $t$ isn't a neighbor of $x$ then $\{i,x,c,t,b,d\}$ induces $2$ $\overline{P_5}$, a contradiction. Let $x'\in
C-N_C(I)$, recall that $x$ isn't adjacent to $x'$; if $x'$ is not adjacent to $t$, the graph
$G[\{a,t,c,x',x,d\}]$contains two induced bulls, a contradiction, then the vertices of $T$ are all adjacent to the
vertices of $C\setminus N_C(I)$.
\end{preuvedeclaim}

Since $G$ is a prime graph, when $I$ has a neighbor in $C$ it follows that the sets $T$ and $I\setminus N_I(C)$ are
empty or $\{a, b, c, d, e\}\cup C\cup N_I(C)$ would be a non trivial module of $G$. Similarly $C\setminus N_C(I)$
contains at most one vertex and thus $G$ is a {\em bundle of $P_5$'s}, one of the graphs depicted in Figure
\ref{fig:FaisceauDeP5}.

From now on, we assume that $I$ has no neighbor in $C$, moreover we may assume that a vertex of $T$ has  a non-neighbor
in $C$ otherwise the set $T$ would be empty ($G$ is a prime graph) and once again, $G$ would be a bundle of $P_5$'s.

%%%%%%%%%%%%%%%%%%%%%%%%%%%%%%%%%%%%%%%%%
\begin{jmclaim}\label{thm:P5HBSparseAvecP5:Clm3}
There is a unique non-edge $c_0t_0$ such that $c_0\in C$ and $t_0\in T$, $c_0$ is adjacent to all other vertices of
$C$, $t_0$ is adjacent to all other vertices of $T$ and has no neighbor in $I$.
\end{jmclaim}
\begin{preuvedeclaim}
Observe first that a vertex of $T$ cannot have two non-neighbors in $T$, otherwise a such vertex say $t$ together with
two non-neighbors in $C$, say $c_1$ and $c_2$ and the vertices $a$, $c$ and $d$ would induce two bulls, a
contradiction. Similarly, a vertex of $C$, say $x$ cannot have two non-neighbors $t_1$ and $t_2$ in $T$ or two bulls
would be induced with the vertices $x$, $c$, $d$, $a$, $t_1$ and $t_2$, a contradiction. If there is two non-edges
$c_1t_1$ and $c_2t_2$ such that $c_1,c_2\in C$ and $t_1,t_2\in T$, those vertices together with $a$ and $e$ would
induce two $P_5$'s or two bulls or two $\overline{P_5}$'s or two $C_5$'s according to the connections between $c_1$ and
$c_2$ and between $t_1$ and $t_2$, a contradiction. If $t_0$ would have a non-neighbor in $T$, say $t$, the vertices
$c_0$, $c$, $t_0$, $t$, $a$ and $e$ would induce two $\overline{P_5}$, a contradiction. A neighbor $i$ of $t_0$ in $I$
together with $c_0$ and the vertices $b$, $c$ and $d$ would induce two bulls in $G$, a contradiction. If $c_0$ is
independent of another member of $C$ say $x$, the graph induced by the vertices $x_0$, $x$, $t_0$, $c$ and $a$ induces
a  bull, as well as $G[\{x_0, x, t_0, c, e\}]$, a contradiction.
\end{preuvedeclaim}

No vertex of $I\cup T\setminus\{t_0\}$ can distinguish the vertices of $\{a, b, c, d, e\}\cup C\cup \{t_0\}$,
consequently $I\cup T\setminus\{t_0\}=\emptyset$. Moreover, $C\setminus\{c_0\}$ contains at most one vertex, it follows
that $G$ has either $7$ or $8$ vertices according to the fact that $C\setminus \{c_0\}$ is empty or not and is
isomorphic to a graph depicted in Figure \ref{fig:FaisceauDeP5}.
\end{preuve}

\begin{figure*}[!ht]
\begin{center}
\input{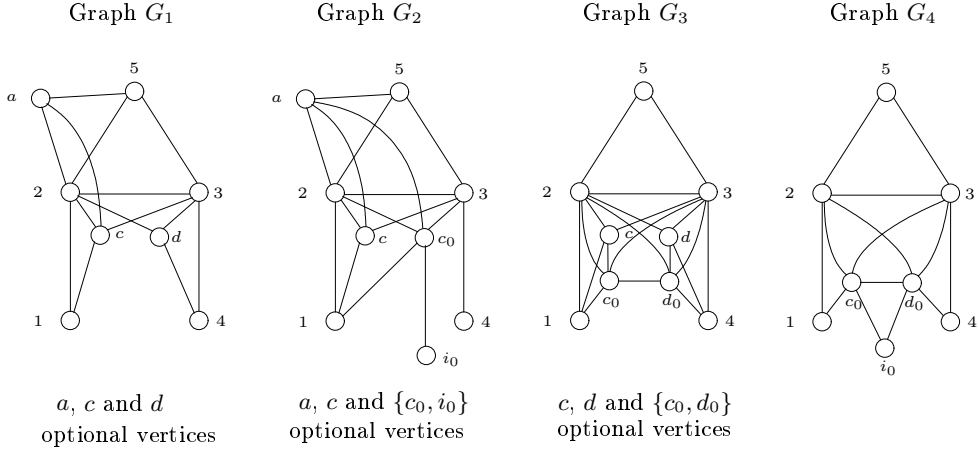}
\caption{The $4$ types of prime  $(P_5,\overline{P_5},Bull)$-sparse graphs which are $(P_5,\overline{P_5},C_5$)-free and contain a $Bull$. } \label{fig:G2G3G4}
\end{center}
\end{figure*}

\begin{theoreme}\label{thm:P5HBSparseAvecBull}
Let $G$ be a prime $(P_5,\overline{P_5},C_5)$-free graph which contains an induced bull. \\
$G$ is $(P_5,\overline{P_5},Bull)$-sparse if and only if $G$ or $\overline{G}$ is isomorphic to one of the graphs depicted in Figure \ref{fig:G2G3G4}.
\end{theoreme}
\begin{preuve}
It is easy to check that all graphs in Figure \ref{fig:G2G3G4} are $(P_5,\overline{P_5},Bull)$-sparse.

Let's consider an induced bull in $G$ whose vertices are numbered $1$, $2$, $3$, $4$, $5$ in such a way that $\{1,2,3,4\}$ induces a $P_4$ whose endpoints are $1$ and $4$ and $5$ is precisely adjacent to $2$ and $3$ and not to $1$ nor $4$.

We consider the $6$ following subsets of $V\setminus \{1, 2, 3, 4, 5\}$.

Let $T$ be the set of vertices which are adjacent to all the members of $\{1, 2, 3, 4, 5\}$ and $I$ be the set of vertices having no neighbor among $\{1, 2, 3, 4, 5\}$.
Let $A$ be the set of vertices being adjacent to $2$ and $5$ and independent of $1$, $3$ and $4$, while $B$ denotes the set of vertices which are adjacent to $3$ and to $5$ and independent of $1$, $2$, and $4$.
Let $C$ be set of vertices which are adjacent to $1$, $2$ and $3$ and independent of $4$ and $5$. $D$ denotes the set of vertices being adjacent to $2$, $3$ and $4$ and independent of $1$ and $5$.

It is easy to see that a vertex $x$ which is partial with respect to $\{1, 2, 3, 4,5\}$ must belong to $A\cup B\cup C \cup D$, in other cases of adjacency the subgraph induced by $\{1, 2, 3, 4, 5,x\}$ would not be $(P_5,\overline{P_5}, Bull)$-sparse. Consequently $V(G)=\{1, 2, 3, 4, 5\}\cup T\cup I\cup A\cup B\cup C\cup D$.

\begin{jmclaim}\label{clm1:thm:P5HBSparseAvecBull}
$C$ is total for $A$ and $T$, $C$ is independent of $B$.
\end{jmclaim}
\begin{preuvedeclaim}
Let $c\in C$.
When  $c$ has a non-neighbor in $A$, say $a$, the set $\{a, 5, 3, c, 1\}$ induces a $P_5$, a contradiction since $G$ is assumed to be $P_5$-free. The vertex $c$ cannot have a neighbor in $B$, or this neighbor together with $c$, $1$, $2$, and $5$ would induce a $\overline{P_5}$, a contradiction.
When  $c$ has a non-neighbor $t$ in $T$, $43ct1$ is a $\overline{P_5}$, a contradiction.
\end{preuvedeclaim}
Let $f$ be an edge preserving mapping such that $f(1)=4$, $f(4)=1$, $f(2)=3$, $f(3)=2$ and $f(5)=5$, we have $f(A)=B$, $f(B)=A$, $f(C)=D$, $f(D)=C$ while $f(T)=T$ and $f(I)=I$. It follows that we can derive from Claims \ref{clm1:thm:P5HBSparseAvecBull} to Claim \ref{clm7:thm:P5HBSparseAvecBull} below many analogous results by considering the mapping $f$ and/or the complementary graph of $G$.
For example the assertion  {\em $C$ is total for $A$} becomes  {\em $D$ is total for $B$} when considering the mapping $f$, while {\em $C$ is total for $T$} becomes {\em  $B$ is independent of $I$} when applied in $\overline{G}$ and  {\em $A$ is independent of $I$} when considering the mapping $f$ in $\overline{G}$.

Let's now examine the connections between vertices of $C$ and $D$ and between vertices of $C$ and $I$.
\begin{jmclaim}\label{clm2:thm:P5HBSparseAvecBull}
If $A\neq \emptyset$  then there is no edge connecting a vertex of $C$ to a vertex of $D$, nor a vertex of $D$ to a vertex of $I$.
\end{jmclaim}
\begin{preuvedeclaim}
Let $a$ be a vertex of $A$.

Assume that $cd$ is an edge ($c\in C$ and $d\in D$), the vertices $c$, $d$, $3$, $5$, $a$ induce a $\overline{P_5}$, a contradiction.

Suppose that $d\in D$ has a neighbor $i$ in $I$, then $id35a$ is a $P_5$, a contradiction
\end{preuvedeclaim}
\begin{jmclaim}\label{clm3:thm:P5HBSparseAvecBull}
If a vertex of $C$ has a neighbor in $I$ then the vertices of $N_I(C)$ are isolated in $I$, the vertices of $N_C(I)$
are isolated in $C$, $T$ is total for $A\cup N_I(C)$, in addition there is a unique edge $c_0i_0$ connecting a vertex
of $C$ to a vertex of $I$ and $i_0$ is isolated in $I$..
\end{jmclaim}
\begin{preuvedeclaim}
Let $c\in C$ and $i\in I$ be adjacent vertices.

If $i$ has a neighbor in $I$, say $i'$, $i'ic34$ is a $P_5$ when $i'$ is independent of $c$ while $\{i, i', c, 2, 3,
4\}$ induces $2$ bulls when $i$ is adjacent to $c$, a contradiction.

If $c$ has a neighbor in $C$, say $c'$, the vertices $i$, $c$, $c'$, $2$, $3$, $4$ induces two bulls, a contradiction.

Let $at$ ($a\in A$, $t\in T$) be a non edge of $G$, then $ict5a$ is a $P_5$ of $G$, a contradiction.

Moreover, observe that a vertex of $C$ cannot have  two neighbors $i$ and $i'$ in $I$, otherwise the vertices $c$, $i$,
$i'$, $2$, $3$, $4$ would induce two bulls, a contradiction. On the same manner, a vertex in $I$ cannot have two
neighbors in $C$, say $c$ and $c'$ or once again two bulls are induced in $G[\{i, c, c', 2, 3, 4\}]$, contradiction.
Consequently, according to Claim \ref{clm3:thm:P5HBSparseAvecBull}, two edges connecting vertices of $C$ to vertices of
$I$ would induce a $2K_2$ and thus this $2K_2$ together with the vertex $2$ would induce a $P_5$ in $G$, a
contradiction.

Let $c_0i_0$ be the unique edge connecting a vertex of $C$  to a vertex of $I$, if $i_0$ has a neighbor, say $i'$ in
$I$, $i'i_0c_034$ would be $P_5$ of $G$, a contradiction.
\end{preuvedeclaim}
\ignore{
\begin{jmclaim}\label{clm4:thm:P5HBSparseAvecBull}
If a vertex of $C$ has a neighbor in $I$ then  there is a unique edge $c_0i_0$ connecting a vertex of $C$ to a vertex
of $I$ and $i_0$ is isolated in $I$.
\end{jmclaim}
\begin{preuvedeclaim}
Observe first that a vertex of $C$ cannot have  two neighbors $i$ and $i'$ in $I$, otherwise the vertices $c$, $i$,
$i'$, $2$, $3$, $4$ would induce two bulls, a contradiction. On the same manner, a vertex in $I$ cannot have two
neighbors in $C$, say $c$ and $c'$ or once again two bulls are induced in $G[\{i, c, c', 2, 3, 4\}]$, contradiction.
Consequently, according to Claim \ref{clm3:thm:P5HBSparseAvecBull}, two edges connecting vertices of $C$ to vertices of
$I$ would induce a $2K_2$ and thus this $2K_2$ together with the vertex $2$ would induce a $P_5$ in $G$, a
contradiction.

Let $c_0i_0$ be the unique edge connecting a vertex of $C$  to a vertex of $I$, if $i_0$ has a neighbor, say $i'$ in
$I$, $i'i_0c_034$ would be $P_5$ of $G$, a contradiction.
\end{preuvedeclaim}
}
\begin{jmclaim}\label{clm5:thm:P5HBSparseAvecBull}
If $C$ has a neighbor in $D$ then $N_D(C)$ is universal in $D$, there is a unique edge connecting a vertex of $C$ to a
vertex of $D$, the vertices of $I$ do not distinguish $c_0$ from $d_0$ and $D\setminus\{d_0\}$ is independent of $I$.
\end{jmclaim}
\begin{preuvedeclaim}
Assume that a vertex $c\in C$ has two neighbors in $D$, namely $d$ and $d'$. In this case the graph induced by the
vertices $c$, $d$, $d'$, $1$, $3$, $5$ contains two bulls, a contradiction. Symmetrically,  a member of $D$ cannot have
two neighbors in $C$.

Moreover, if $d$ is independent of some other vertex  of $D$, namely  $d'$, the set $\{c, 3, d, 1, 5, d'\}$ induces two
bulls, a contradiction, thus $N_D(C)$ is universal in $D$, similarly $N_C(D)$ is universal in $C$.

If $cd$ and $c'd'$ ($c,c'\in C$, $d,d'\in D$) are two distinct edges, $G[\{c, c', d, d', 4\}]$ is a $\overline{P_5}$, a
contradiction which proves the uniqueness of an edge connecting $C$ to $D$.

Assume without loss  of generality that $i\in I$ is adjacent to $c_0$ and not to $d_0$. the subgraph induced by $1$,
$i$, $c_0$, $d_0$, $3$ and $5$ contains two bulls, a contradiction.

Finally, suppose that  $d\in D$, distinct from $d_0$ is adjacent to $i\in I$, then $G$ contains a $P_5$ ($idd_035$ if
$i$ is adjacent to $d_0$ and $dic_025$ if $i$ is  not adjacent to $d_0$), a contradiction.
\end{preuvedeclaim}
\ignore{
\begin{jmclaim}\label{clm6:thm:P5HBSparseAvecBull}
Let  $c_0d_0$ ($c_0\in C, d_0\in D$) be the unique edge connecting a vertex of $C$ to a vertex of $D$,  the vertices of
$I$ do not distinguish $c_0$ from $d_0$ and $D\setminus\{d_0\}$ is independent of $I$.
\end{jmclaim}
\begin{preuvedeclaim}
Assume without loss  of generality that $i\in I$ is adjacent to $c_0$ and not to $d_0$. the subgraph induced by $1$,
$i$, $c_0$, $d_0$, $3$ and $5$ contains two bulls, a contradiction.

Suppose now that  $d\in D$, distinct from $d_0$ is adjacent to $i\in I$, then $G$ contains a $P_5$ ($idd_035$ if $i$ is
adjacent to $d_0$ and $dic_025$ if $i$ is  not adjacent to $d_0$), a contradiction.
\end{preuvedeclaim}
}
\begin{jmclaim}\label{clm7:thm:P5HBSparseAvecBull}
At least one of the sets $A$, $B$, $C$, $D$ is empty.
\end{jmclaim}
\begin{preuvedeclaim}
Let $a\in A$, $b\in B$, $c\in C$, $d\in D$. We know by Claim \ref{clm1:thm:P5HBSparseAvecBull} that $a$ is connected to $c$ and not to $d$ and that $b$ is connected to $d$ and not to $c$, Claim\ref{clm2:thm:P5HBSparseAvecBull} asserts that $c$  is not adjacent to $d$ while  $a$ and $b$ are connected. Consequently $1cabd$ is a $P_5$, a contradiction.
\end{preuvedeclaim}
According to Claim \ref{clm7:thm:P5HBSparseAvecBull} we will now discuss on the number of empty sets among $A$, $B$, $C$ and $D$ and prove that $G$  or $\overline{G}$ is isomorphic to one of the graphs depicted in Figure \ref{fig:G2G3G4}.
\paragraph{Case $1$~: The sets $A$, $B$, $C$ and $D$ are all empty.}

Recall that $G$ is prime, thus the sets $T$ and $I$ are also empty, for otherwise $\{ 1, 2, 3, 4, 5\}$ would be a
non-trivial module. Consequently $G$ is a bull, a graph isomorphic to $G_1$ in Figure \ref{fig:G2G3G4} when $a$, $c$
and $d$ are missing.
\paragraph{Case $2$~: Three of the sets $A$, $B$, $C$ and $D$ are empty.}

Assume without loss of generality that $C\neq\emptyset$. We know by Claim \ref{clm1:thm:P5HBSparseAvecBull} that $C$ is
total for  $T$

If $C$ has no neighbor in $I$, no vertex of $T\cup I$ can distinguish the members of $\{1, 2, 3, 4, 5\}\cup C$ and by the primality of $G$ the sets $T$ and $I$ are empty while $C$ is reduced to a single vertex. In this case $G$ is isomorphic to $G_1$ where $a$ and $d$ are missing.

If $C$ has a neighbor in $I$ we know by Claim \ref{clm3:thm:P5HBSparseAvecBull} that there is a unique edge, namely
$c_0i_0$ connecting $C$ to $I$. We consider the following decomposition of $C$ and $I$~: $C=\{c_0\}\cup(C\setminus
\{c_0\})$, $I=\{i_0\}\cup(I\setminus \{i_0\})$.

\noindent By construction $C\setminus\{c_0\}$ is independent of $I$ and $I\setminus\{i_0\}$ is independent of $C$ while $i_0$ has no neighbor in $I\setminus\{i_0\}$  and  $c_0$ has no neighbor in $C\setminus\{c_0\}$ (Claim \ref{clm3:thm:P5HBSparseAvecBull}). Moreover  $i_0$ is completely adjacent to $T$ (Claim \ref{clm3:thm:P5HBSparseAvecBull}).

\noindent Consequently $T\cup (I\setminus\{i_0\})=\emptyset$ or  the set $\{1, 2, 3, 4, 5, c_0, i_0\}\cup  (C\setminus\{c_0\})$ would be a non trivial module of $G$, a contradiction. In addition $C\setminus\{c_0\}$ is either a singleton, say $\{c\}$ or empty and  $G$ is isomorphic to  $G_2$ without the vertex $a$ and where  $c$ is possibly missing if $C\setminus\{c_0\}=\emptyset$ (see Figure \ref{fig:G2G3G4}).
\paragraph{Case $3$~: Among $A$, $B$, $C$ and $D$ exactly two sets are empty.}

\noindent  Due to symmetries we only consider three different situations.

Let first suppose that $B=C=\emptyset$.

\noindent We know (Claim \ref{clm1:thm:P5HBSparseAvecBull}) that $A$ and $D$ are independent, $D$ is total for $T$ and $A$ is independent of $I$. Moreover $D$ is independent of $I$ (Claim \ref{clm2:thm:P5HBSparseAvecBull}) and thus $A$ is total for $T$. Because of the primality of $G$ the set $T\cup I$ is empty and $A$ as well as $D$ is a singleton. Consequently $G$ is isomorphic to $G_2$ without the vertex $c$ (Figure \ref{fig:G2G3G4}).

Assume in a second stage  that $B=D=\emptyset$.

\noindent We know by Claim \ref{clm1:thm:P5HBSparseAvecBull} that $C$ is total for  $T$ and $A$ is independent of $I$.

If there is an edge between $C$ and $I$, it is unique (Claim \ref{clm3:thm:P5HBSparseAvecBull}), let's denote this edge
$c_0i_0$. In this case $A$ is totally adjacent to $T$ (Claim \ref{clm3:thm:P5HBSparseAvecBull}), the set
$C\setminus\{c_0\}$ is independent of $I$ and by construction $c_0$ is independent of $I\setminus\{i_0\}$, $i_0$ is
independent of  $I\setminus\{i_0\}$  and  $c_0$ has no neighbor in $C\setminus\{c_0\}$ (Claim
\ref{clm3:thm:P5HBSparseAvecBull} again). It follows that the prime graph $G$ is isomorphic to $G_2$ (Figure
\ref{fig:G2G3G4}) where $c$ can miss if $C\setminus \{c_0\}$ is empty.

If there is no  connection between $C$ and $I$, some vertex of $A$ can have a non neighbor in $T$, we are then in a similar situation than above in the complementary graph of $G$.

When $C$ is independent of $I$ and $A$ is total for $T$ the graph is isomorphic to graph $G_1$ in Figure \ref{fig:G2G3G4}.

Finally let's study the case $A=B=\emptyset$.

\noindent We know that $C$ and $D$ are totally adjacent to $T$ (Claim \ref{clm1:thm:P5HBSparseAvecBull}).

If $C$ and $D$ are not connected it is easy to see that $C$ and $D$ are not adjacent to $I$. As a matter of fact, suppose on the contrary that $c_0i_0$ is an edge ($c_0\in C$ and $i_0\in I$) and that $d$ is some vertex in $D$. If $d$ and $i_0$ are not connected, $i_0c_02d4$ is a $P_5$ and $i_0c_03d4$ is a $\overline{P_5}$ if $d$ and $i_0$ are adjacent, a contradiction in both cases.
Consequently, $G$ being prime is isomorphic to the graph $G_1$ in Figure \ref{fig:G2G3G4} where $d$ misses.

When $C$ has a neighbor in $D$, we consider the unique edge connecting $C$ to $D$, namely $c_0d_0$ ($c_0\in C, d_0\in
D$). By Claim \ref{clm5:thm:P5HBSparseAvecBull}, $c_0$ is universal in $C$ and $d_0$ is universal in $D$. We know
(Claim \ref{clm5:thm:P5HBSparseAvecBull}) that only $c_0$ and $d_0$ can have a neighbor in $I$.

If it is not the case $G$ is isomorphic to $G_3$ in Figure \ref{fig:G2G3G4} without $c$ or $d$ if  $C\setminus\{c_0\}$ or  $D\setminus\{d_0\}$ is empty.
If, on the contrary, $c_0$ and $d_0$ have a neighbor, say $i_o$ in $I$, $C=\{c_0\}$ (or $\{1, c, c_0, i_0, d_0, 4\}$ where $c$ is a vertex of $C$ distinct from $c_0$ induces two bulls, a contradiction) and similarly $D=\{d_0\}$. Consequently $G$ is isomorphic to the graph $G_4$ of Figure \ref{fig:G2G3G4}.
\paragraph{Case $4$~: Among $A$, $B$, $C$ and $D$ exactly one set is empty.}

\noindent For convenience we will suppose that $B=\emptyset$.

By Claim \ref{clm1:thm:P5HBSparseAvecBull}, $A$ is completely adjacent to $C$ and independent of $D$. There is no edge connecting a vertex of $C$ to a vertex of $D$ (Claim \ref{clm2:thm:P5HBSparseAvecBull}).

Moreover $C$ and $D$ are completely adjacent to $T$ and $A$ is independent of $I$ (Claim \ref{clm1:thm:P5HBSparseAvecBull}). In addition, there is no connection between $D$ and $I$ (Claim \ref{clm2:thm:P5HBSparseAvecBull}) and similarly $A$ is total for $T$.

If there is no edge between $C$ and $I$, the sets $T$ and $I$ must be empty (or $\{1, 2, 3, 4, 5\}\cup A\cup C\cup D$
would be a non trivial module of $G$) and $A$, $C$, $D$ are singletons. In this case $G$ is isomorphic to $G_2$ in
Figure \ref{fig:G2G3G4}.

When there is a unique edge $c_0i_0$ between $C$ and  $I$ ($c_0\in C, i_0\in I$), once again $I\setminus\{i_0\}$ is completely independent of $C\cup \{i_0\}$ while $\{c_0,i_0\}$ has no connections with $C\setminus\{c_0\}$ (Claim \ref{clm3:thm:P5HBSparseAvecBull}). Consequently, $G$ is isomorphic to $G_2$ where $c$ misses if $C=\{c_0\}$.
 \end{preuve}
It follows from Theorem \ref{thm:StructureDesP5HB-Free},  \ref{thm:LesC5DansLesP5H-Sparse},
\ref{thm:StructureDesP5HB-Free} and \ref{thm:P5HBSparseAvecBull} that a prime $(P_5,\overline{P_5},Bull)$-sparse graph
or its complement is either a $C_5$ or a $P_5$-free bipartite graph or a bundle of $P_5$'s (see Figure
\ref{fig:FaisceauDeP5}) or is a graph on less than $10$ vertices. This leads to a linear time recognition algorithm for $(P_5,\overline{P_5},Bull)$-sparse graphs, moreover those graphs have
bounded clique-width (see \cite{CouMakRot98}).

\bibliographystyle{plain}
\bibliography{P5-Sparse}
\end{document}